\documentclass[12pt]{article}

\usepackage{cite}
\usepackage{epsfig}
\usepackage{amsmath}
\usepackage{booktabs}

\newcommand{\anc}{\rule{0mm}{0mm}}

\newcommand{\Eslash}{{\not{\!\!E}}}

\newcommand{\st}{{\tilde{\tau}}}

\newcommand{\sq}{{\tilde{q}}}
\newcommand{\sqR}{{\tilde{q}_{\rm R}}}
\newcommand{\sqL}{{\tilde{q}_{\rm L}}}
\newcommand{\su}{{\tilde{u}}}
\newcommand{\suR}{{\tilde{u}_{\rm R}}}
\newcommand{\suL}{{\tilde{u}_{\rm L}}}
\newcommand{\sd}{{\tilde{d}}}

\newcommand{\sdL}{{\tilde{d}_{\rm L}}}

\newcommand{\ssL}{{\tilde{s}_{\rm L}}}

\newcommand{\scL}{{\tilde{c}_{\rm L}}}

\newcommand{\sF}{{\tilde{f}}}
\newcommand{\go}{\tilde{g}}

\newcommand{\cha}{\tilde{\chi}}
\newcommand{\neu}{\tilde{\chi}^0}
\newcommand{\mcha}[1]{m_{\tilde{\chi}^\pm_{#1}}}
\newcommand{\mneu}[1]{m_{\tilde{\chi}^0_{#1}}}

\newcommand{\gev}{{\rm \ GeV}}

\newcommand{\mycaption}[1]{\caption{\sl #1}}

\begin{document}
\thispagestyle{empty}

\def\thefootnote{\fnsymbol{footnote}}

\begin{flushright}
FERMILAB--Pub--08/15--T\\
QUIGLEYS--Pub--43--Jefferson\\
ZH--TH 13/06
\end{flushright}

\vspace{1cm}

\begin{center}

{\Large\sc {\bf Determining the SUSY-QCD Yukawa coupling}}
\\[3.5em]
{\large\sc 
A.~Freitas$^{1}$
and
P.~Skands$^{2}$
}

\vspace*{1cm}

{\sl
$^1$ Institut f\"ur Theoretische Physik,
        Universit\"at Z\"urich, \\ Winterthurerstrasse 190, CH-8057
        Z\"urich, Switzerland

\vspace*{0.4cm}

$^2$ Theoretical Physics, Fermi National Accelerator Laboratory, P.\
O.\ Box 500, Batavia, IL-60510, USA 

\vspace*{0.4cm}

}

\end{center}

\vspace*{2.5cm}

\begin{abstract}
Among the firm predictions of softly broken supersymmetry is 
the identity of gauge couplings and the corresponding Yukawa couplings
between gauginos, sfermions and fermions. 
In the event that a SUSY-like spectrum of new particles is discovered
at future colliders, a key follow-up 
will be to test these relations experimentally. 
In detailed studies it has been found that the
SUSY-Yukawa couplings of the electroweak sector can be studied with great
precision at the ILC, but a similar analysis for the Yukawa coupling of
the SUSY-QCD sector is far more challenging. Here a first
phenomenological study for determining this coupling is presented, using a
method which combines information from LHC and ILC.
\end{abstract}

\def\thefootnote{\arabic{footnote}}
\setcounter{page}{0}
\setcounter{footnote}{0}

\newpage


\section{Introduction}

With more than 1 fb$^{-1}$ of luminosity on tape at the Tevatron
experiments, and with the LHC and ILC on the horizon, 
TeV scale physics is fast entering the realm of experimental
study. 
Among the most interesting and comprehensively studied 
possibilities for observable new physics is 
supersymmetry (SUSY). Interesting in its own right by virtue 
of being the largest possible space-time symmetry \cite{haag75}, 
it was chiefly with a number of additional theoretical 
successes in the early eighties that supersymmetry gained widespread
acceptance, among these successful 
gauge coupling unification, radiative breaking
of electroweak symmetry, and a natural candidate for Dark Matter.

Stated briefly, supersymmetry promotes all the fields of the Standard
Model (SM) to superfields, all its multiplets to supermultiplets. Each
such multiplet contains one boson and one fermion, which are 
related to each other by supersymmetry. Hence not only should all the
SM particles (including at least one more Higgs doublet) have
`sparticle' superpartners, with spins
differing by 1/2, but also the interactions of these superpartners
should be fixed by supersymmetry. Even when considering broken
supersymmetry (the only phenomenologically viable option), many of
these relations persist, or are only modified to a tractable degree. 
Hence, part of the effort of testing the hypothesis, should suitable
particles be discovered, will be to check whether their interactions
do obey the supersymmetric relations or not. It is the feasibility of
a particular such study we shall address here. 

Among the most intuitively clear relations is that the supersymmetric 
partner of a gauge boson, a gaugino, must couple with the same
strength to gauge charge as its partner does. That is, the Yukawa
coupling $\hat{g}$ between a gaugino interacting with a fermion and a sfermion
must be identical to the corresponding SM gauge coupling $g$ between a
gauge boson and two (s)fermions. More specifically, 
denoting an SM fermion (gauge boson) by $f$ ($V$), and its sfermion
(gaugino) superpartner by $\tilde{f}$ ($\tilde{V}$), 
\begin{align}
g &\equiv g(Vff) = g(V\sF\sF), &
\hat{g} &\equiv g(\tilde{V} f \sF)~, 
\end{align}
with 
\begin{equation}
g = \hat{g}
\end{equation}
required by supersymmetry. 
Indeed this relation is vital for the cancellation of quadratic
divergencies in the radiative corrections to the Higgs mass. 

In what follows, we now constrain our attention to the Minimal
Supersymmetric Extension of the Standard Model (MSSM), with $CP$ and
$R$-parity conserved. Hence the Lightest Supersymmetric Particle
(LSP), usually the lightest neutralino, is stable, and supersymmetric
particles can only be produced in pairs.

If the supersymmetric particles are light enough to be produced at the next
generation of colliders, it has been shown \cite{susyid} that 
the supersymmetric Yukawa couplings in the electroweak
sector can be precisely tested at a high-energy $e^+e^-$ collider.
The $t$-channel contributions to neutralino/chargino \cite{ckmz} and slepton
production \cite{susyid,slepo,slep} in $e^+e^-$ and $e^-e^-$ collisions are
directly dependent on the U(1) and SU(2) Yukawa couplings. From measurements of
total and differential cross sections, these couplings can thus 
be extracted with a precision at the per-cent level \cite{susyid,slepo,slep}.

However, the analysis of the supersymmetric Yukawa coupling $\hat{g}_{\rm s}$ 
in the SU(3) QCD sector is much more difficult. 
At the ILC this interaction can be studied in
the process $e^+e^- \to q \sq^* \go$, $\bar{q} \sq \go$
\cite{BMWZ}. While this approach is in principle straightforward, it 
suffers both from a very small cross section, at most 
about 1 fb for squark and gluino masses of a few hundred GeV
\cite{BMWZ}, and also from the daunting challenge of 
selecting  the signal from much larger $t\bar{t}$ and squark backgrounds.
At the LHC on the other hand, squarks and gluinos with masses below
2--3 TeV are copiously produced, and their pair production cross sections
depend directly on the supersymmetric Yukawa coupling $\hat{g}_{\rm
  s}$. However, measurements of total cross sections are
exceedingly difficult in this environment, with 
typically only one or two specific decay channels of the squarks and
gluinos experimentally accessible \cite{lhclc}. 

Here we consider a combination --- the relevant branching ratios are to be
determined at an $e^+e^-$ collider and combined with exclusive
cross section measurements in selected channels at the LHC. Together, 
this information determines the total squark or gluino production
cross section, from  which a value for the SUSY-QCD Yukawa coupling
$\hat{g}_{\rm s}$ can be extracted.

In this publication, a first analysis is presented to explore whether such a
coherent LHC/ILC analysis can provide a useful determination of the SU(3)
Yukawa coupling. Since it is not clear a priori that the goal is achievable
at all, we constrain our attention to a rather optimistic scenario in
this study, with reasonably low masses, large cross sections,
and a fairly large squark-gluino mass splitting. 

In the next section, the production of squarks at the LHC is studied
and the basic phenomenology laid out. 
The relevant branching ratios of the squarks and how they 
could be determined from
squark pair production at $e^+e^-$ colliders is the topic of section
\ref{sc:ilc}. We round off by presenting the combined result (in the
particular scenario studied here) and give a few concluding remarks.


\section{Squark production at the LHC}
\label{sc:lhc}
\subsection{Phenomenology and Strategy -- LHC}
\begin{figure}
\parbox[b][6cm][t]{16.5cm}{
\anc\hspace{2mm}%
\psfig{figure=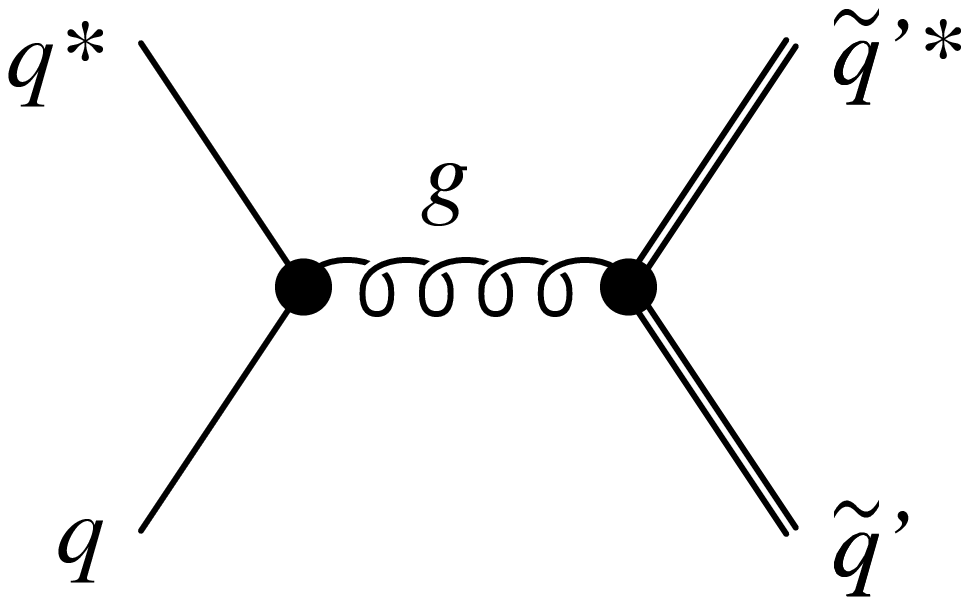, width=5.3cm}
\psfig{figure=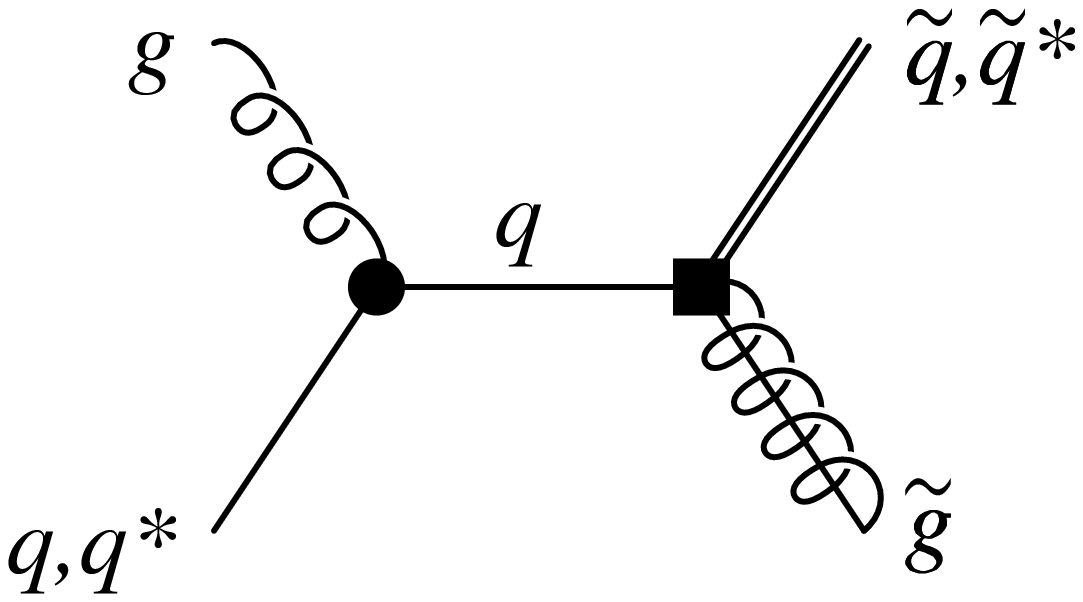, width=6.3cm}
\psfig{figure=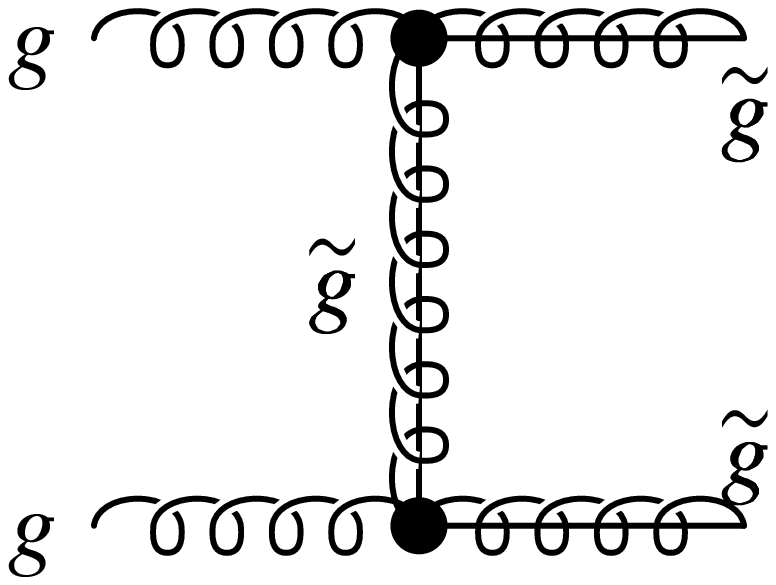, width=4cm}\\
\psfig{figure=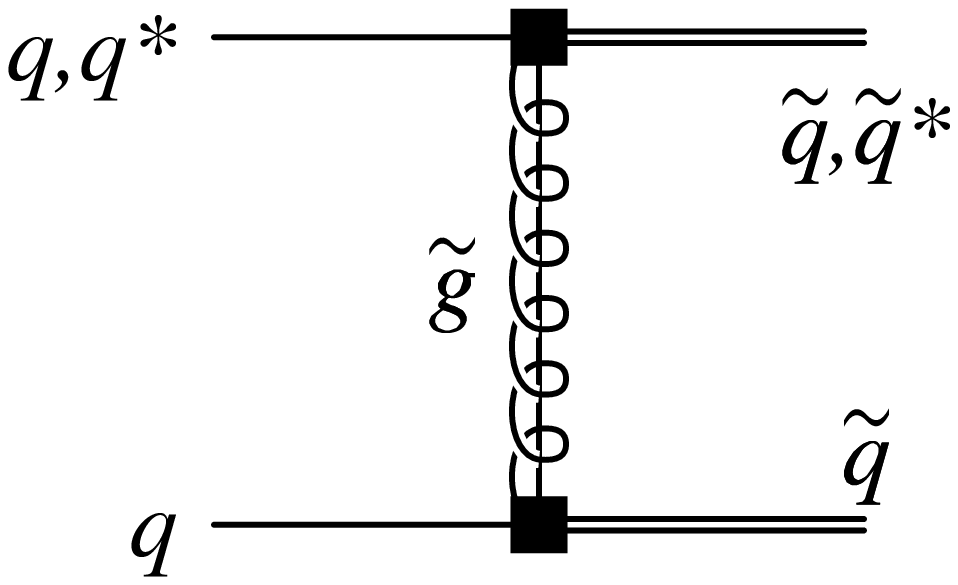, width=5.5cm}
\hspace{7mm}
\raisebox{1mm}{\psfig{figure=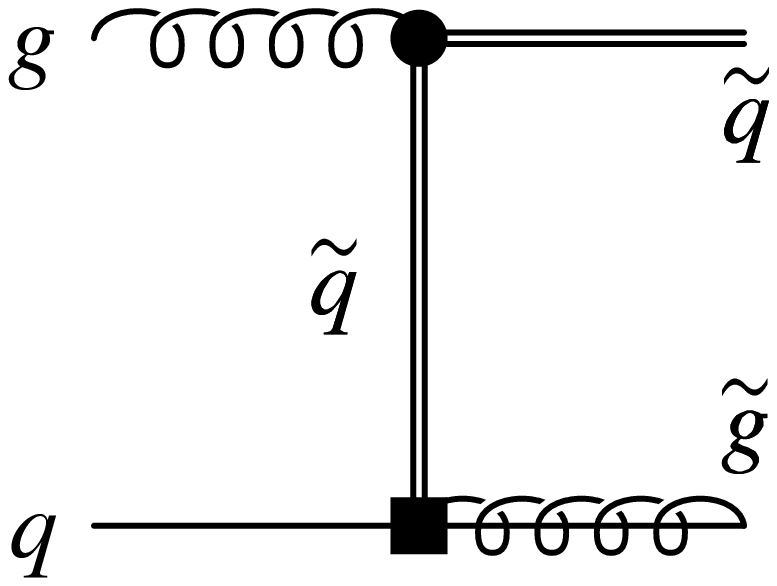, width=3.8cm}}
\hspace{12mm}
\raisebox{1mm}{\psfig{figure=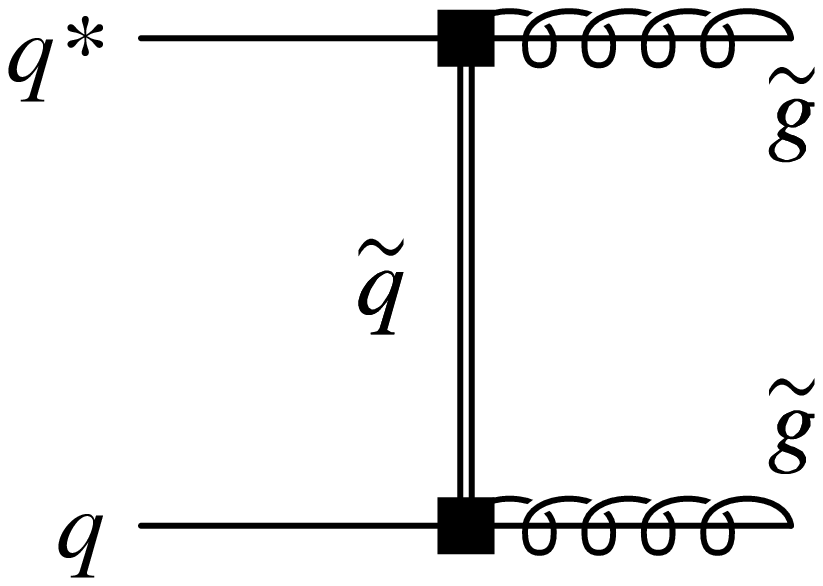, width=4.3cm}}
}
\mycaption{Some examples for Feynman diagrams for 
partonic squark and gluino production in hadron
collisions. Dots indicate the gauge coupling $g_{\rm s}$, while squares stand
for the Yukawa coupling $\hat{g}_{\rm s}$.}
\label{fg:dia1}
\end{figure}

In $pp$ collisions, squark and gluino production occurs in a
variety of combinations, often with both $s$- and 
$t$-channel diagrams as well as several different partonic initial
states contributing to the same final state. Fig.~\ref{fg:dia1}
contains a condensed summary of these, with explicit illustration of
which vertices go with the QCD gauge (dots) and Yukawa (squares)
couplings, respectively. 
Naively, there are several processes from which 
one might attempt to extract the Yukawa coupling. 
However, final states which receive
contributions from both gauge and Yukawa couplings make the analysis more
complicated and model-dependent. To isolate the Yukawa coupling, we note that
the production of same-sign squarks (e.g.\ $\suL\suL$, $\suR\suR$, ...) 
proceeds only through the diagram shown in the lower left corner of
Fig.~\ref{fg:dia1}, if one neglects the much smaller electroweak
contributions. 
Hence, to excellent approximation, this process 
depends solely on the supersymmetric
Yukawa coupling $\hat{g}_{\rm s}$, and to the extent it can be 
isolated from the background, a clean determination of $\hat{g}_{\rm
  s}$ should be possible. Note that, since the final state flavours are
locked to the initial state ones, in $pp$ collisions this process 
dominantly produces $\su$ and $\sd$ squarks, with smaller admixtures of
sea-flavoured squarks, in direct proportion to the quark
content of the proton at the relevant $x$ and $Q^2$ values. 

Due to the flavour locking, only the first two generations of
squarks are thus relevant, for which mixing effects are small and
we can take mass and current eigenstates to be identical to good
approximation. That is, 
the heavier $\sq$ mass eigenstate is pure $\sqL$ (weak isospin
doublet), and the lighter one pure $\sqR$ (weak isospin singlet). 
Nominally, the lighter one would be the better target for a
high-statistics study, simply due to phase space, but since it doesn't
couple to weak interactions, it decays almost exclusively 
via the hypercharge coupling to a same-flavour quark and the
LSP. Since charge tagging for light-flavour jets is exceedingly
difficult, this decay mode
effectively obscures the fact that we had same-flavour squarks to
begin with. Moreover, since it only contains a jet and missing energy,
the mode would be extremely challenging to separate from the background.  
The only feasible avenue thus appears to be to use
flavour/charge tagging modes of the heavier mass eigenstates,
the $\sqL$. 

For $\sqL$, 
the charge of the squark can be tagged through a chargino decay chain,
\begin{align}
\suL &\to d \, \cha^+_1 \to d \, l^+ \, \nu_l \, \neu_1, &
\sdL &\to u \, \cha^-_1 \to u \, l^- \, \bar{\nu}_l \, \neu_1, \\
\suL^* &\to \bar{d} \, \cha^-_1 \to \bar{d} \, l^- \, \bar{\nu}_l \, \neu_1, &
\sdL^* &\to \bar{u} \, \cha^+_1 \to \bar{u} \, l^+ \, \nu_l \, \neu_1,
\end{align}
and similarly for $\ssL$ and $\scL$. 
For a given squark flavor, the sign of the final-state lepton is related to the
charge of the (anti-)squark. The production of same-sign squarks through the
diagram in the lower left corner of Fig.~\ref{fg:dia1} 
with this decay channel will therefore lead to
same-sign leptons in the final state, while other direct squark
production processes  will tend to produce opposite-sign leptons in the final
state. At this level, the signal is thus characterized by two
same-sign leptons, two hard jets and missing transverse energy in the
final state.

A very problematic background
can come from gluino pair and mixed gluino-squark production if $m_{\go} >
m_{\sqL}$. In this case, gluinos can decay into quarks and squarks, $\go
\to q \, \sqL$, generating a component of 
two same-sign squarks plus additional jets which has a much more
complicated (see
Fig.~\ref{fg:dia1}) dependence on both the QCD gauge coupling and the 
Yukawa coupling we are interested in.  This
background becomes particularly challenging if the mass difference 
$m_{\go} - m_{\sqL}$ is small, since then the additional jets from the
gluino decay will be soft. Due to the large overall mass scale in the
process, the bulk of the cross section will contain several 
extraneous QCD jets, e.g.\ from  initial state radiation, which can be
significantly harder than this \cite{susyjets}, and hence the separation of
squark production into direct and gluino-induced samples 
will be extremely non-trivial.

In the following we will only consider a scenario where the $m_{\go} -
m_{\sqL}$ mass difference is sufficiently large to allow a veto on
additional jets for gluino background reduction. Although this 
will also reduce the signal somewhat, there are indications 
\cite{susyjets} that 
the particular signal process considered here, being dominated by valence
quarks in the initial state, is associated with less extra radiation
than other SUSY processes, which generally (in $pp$) have one or more 
gluons and/or sea quarks as initiators.  

The most important backgrounds from Standard Model sources are 
$W^\pm W^\pm jj$ (same-sign $W$'s), where $j$ is a light-flavour
jet, and semi-leptonic $t\bar{t}$, with the second lepton 
coming from the decay of a bottom quark. Due to the large total
$t\bar{t}$ cross section, this can result in a sizable background. 
Below we include both these sources in our estimates, 
though the level of sophistication is obviously not 
anywhere near what would be required for a real experimental analysis. 

\subsection{Numerical Results -- LHC}
For the numerical analysis we use the scenario given in the appendix.
This scenario is similar to the Snowmass point SPS1a \cite{sps}, with the
exception that the gluino mass is raised to 700 GeV. Since $\tan\beta
= 10$, the lightest scalar tau eigenstate has a
sizeable left-handed component, and since the other left-handed sleptons
are too heavy, the chargino mainly decays into scalar taus, 
which subsequently decay into taus. 
To trace the charge explicitly, we here restrict our attention only to 
the leptonic tau branching fraction. The decay chain for the signal
process is then 
\begin{eqnarray}
\suL & \stackrel{65\%}{-\!\!\!-\!\!\!\longrightarrow} & d \, \cha^+_1
\stackrel{100\%}{-\!\!\!-\!\!\!\longrightarrow} d \, \tau^+ \, \nu_\tau \,
\neu_1 \stackrel{35\%}{-\!\!\!-\!\!\!\longrightarrow} d \, \ell^+ + \Eslash,\nonumber\\
\sdL & \stackrel{61\%}{-\!\!\!-\!\!\!\longrightarrow} & u \, \cha^-_1
\stackrel{100\%}{-\!\!\!-\!\!\!\longrightarrow} u \, \tau^- \, \bar{\nu}_\tau \,
\neu_1 \stackrel{35\%}{-\!\!\!-\!\!\!\longrightarrow} u \, \ell^- + \Eslash,
\qquad \ell = e, \, \mu. 
\label{eq:dec}
\end{eqnarray}
The numbers above the arrows indicate branching fractions and $\Eslash$
stands for missing energy.

Both signal and top and gluino backgrounds were simulated with {\sc Pythia}
6.326 \cite{pythia}, i.e.\ with the hard $2\to2$ scattering process 
calculated at leading order, dressed up with sequential resonance decays, 
parton showers (virtuality-ordered `power' showers), 
underlying event (`Tune A' \cite{tunea}),  
string hadronisation, and hadron decays (the $\pi^0$ was set 
stable to optimize efficiency).  
The $WWjj$ background was generated with {\sc MadEvent} \cite{mad}, i.e.\ at
leading order with stable outgoing $W$ bosons and no fragmentation. 

The cross sections for squark and gluino production were 
normalized by the $K$-factors obtained with {\sc Prospino 2.0}
\cite{nlo}, which includes 
next-to-leading order QCD corrections. Top production was likewise
normalized to a total inclusive LHC 
cross section of 800~pb, while for the $W^\pm
W^\pm jj$ background only leading order results are available, 
translating to a comparatively larger theoretical uncertainty on this
contribution. We further used {\sc Pythia} to estimate 
the possible contamination from $VV$ ($ZZ$, $W^\pm Z$, and $W^+W^-$)
production, and found it to be small.  

Finally, a mock detector was set up, based on a toy calorimeter
spanning the pseudorapidity range $|\eta|<5$ with a
resolution of $0.1\times 0.1$ in $\eta\times\phi$ space 
and a primitive UA1-like cone jet algorithm, 
with a cone size of $\Delta R=0.4$. Muons and electrons 
were reconstructed inside $|\eta|<2.5$ and an
isolation criterion was further imposed, requiring both less 
than 10 GeV of additional energy deposited in a cone of size
$\Delta R=0.2$ around the lepton and also no reconstructed jets
with $p_{\rm T,j} > 25$ GeV closer than $\Delta R=0.4$ around the
lepton. These criteria duplicate the default settings of the ATLFAST simulation
package \cite{atlas}. 

Based on the general characteristics of the squark signal, we chose a 
set of preselection cuts, to broadly define the signal region:
\begin{itemize} 
\item at least 100 GeV of missing energy.
\item at least 2 jets with $p_{\rm T,j} > 100$ GeV.
\item Exactly two isolated leptons $\ell = e,\mu$ with $p_{\rm T,\ell}
  > 7$ GeV. 
\end{itemize}

As evident from 
Tab~\ref{tab:cuts}, after preselection most backgrounds are still much larger
than the same-sign squark signal. The backgrounds can be further
reduced by making use of the following characteristics.
\begin{table}
\begin{tabular}{llrr|rrrrr}\toprule
\multicolumn{2}{l}{Cross Sections} & Signal & \multicolumn{5}{l}{Backgrounds} \\
\multicolumn{2}{l}{ $\sum_{q=u,d,s,c}\sigma$ (fb)} & $\sqL\sqL$ & \bf Sum &
$t\bar{t}$ & $W^\pm W^\pm jj$ &  $\sqL\go$ & $\sqL\sqL^*$ & $\go\go$
\\\cmidrule{1-9}
\multicolumn{2}{l}{Total} & 2100 & - & 8$\times 10^5$ & - & 7000 & 1350 & 3200
\\\cmidrule{1-9}
\multicolumn{2}{l}{Preselection}    & 49.2 & 384.6 & 177.7 & - & 136.4 & 23.2 & 47.3
\\
\multicolumn{2}{l}{b-veto}          & 17.1 & 31.4 &  13.0 &  - & 10.3 & 7.1 & 1.0 
\\
\multicolumn{2}{l}{$\Eslash > 150$ GeV}&15.1 & 22.2 & 6.1 & - & 9.0 & 6.2 & 0.9 
\\
A) & $p_{T,j_3}<75$ GeV&10.1&9.5& 3.6 & N/A &  2.3 & 3.5 & 0.1     
\\
   & $p_{T,j_1}>200$ GeV&\bf 8.9&\bf $<$7.6& 1.7 & $<$0.7 &  2.0 & 3.2 & 0.1
\\
B) & $p_{T,j_3}<50$ GeV& 7.8&5.9& 2.4 & N/A & 1.0 & 2.5 & 0.03
\\
  & $p_{T,j_1}>200$ GeV& \bf 7.0&\bf $<$4.9& 1.0 & $<$0.7 & 0.8 & 2.3 & 0.03
\\
  & ($Q_\ell = +1$) &    4.8&$<$3.0& 0.6 & $<$0.5 &  0.5 & 1.4 & 0.01\\\bottomrule
\end{tabular}
\mycaption{Total and progressively reduced cross sections estimated from
  the cuts defined in the text, for the signal, summed background, and
  individual backgrounds, respectively. Two different sets of cuts
  have been considered, A and B, the latter incorporating a more
  severe 3rd jet veto. For illustration, the effect of including a
  further cut on the lepton sign is included for set B, though note
  that this last cut is not actually made in the analysis.}
\label{tab:cuts}
\end{table}

\subsubsection{Cuts --- Set A}

The signal is flavour-locked to the initial state and hence
  contains almost no heavy-flavour squarks. On the other hand, both 
$t\bar{t}$ and $\go\go$ backgrounds contain large $b$ fractions 
(it is a general feature of models with universal
high-scale squark masses that gluinos decay dominantly into third-generation
squarks). 

\begin{figure}
\vskip-4mm
\begin{tabular}{cc}
\includegraphics*[scale=0.5]{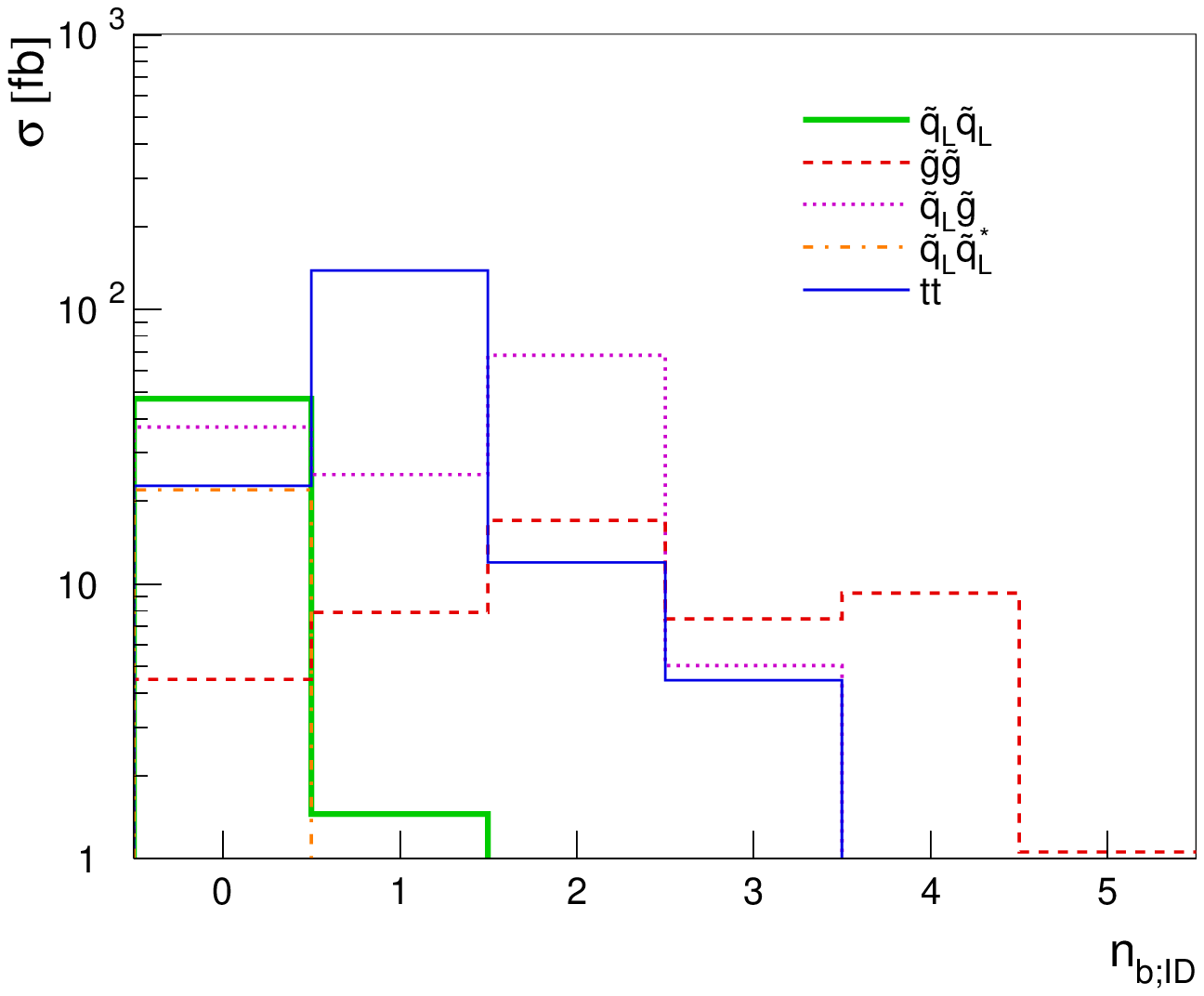}\vspace*{-4mm}
&
\includegraphics*[scale=0.5]{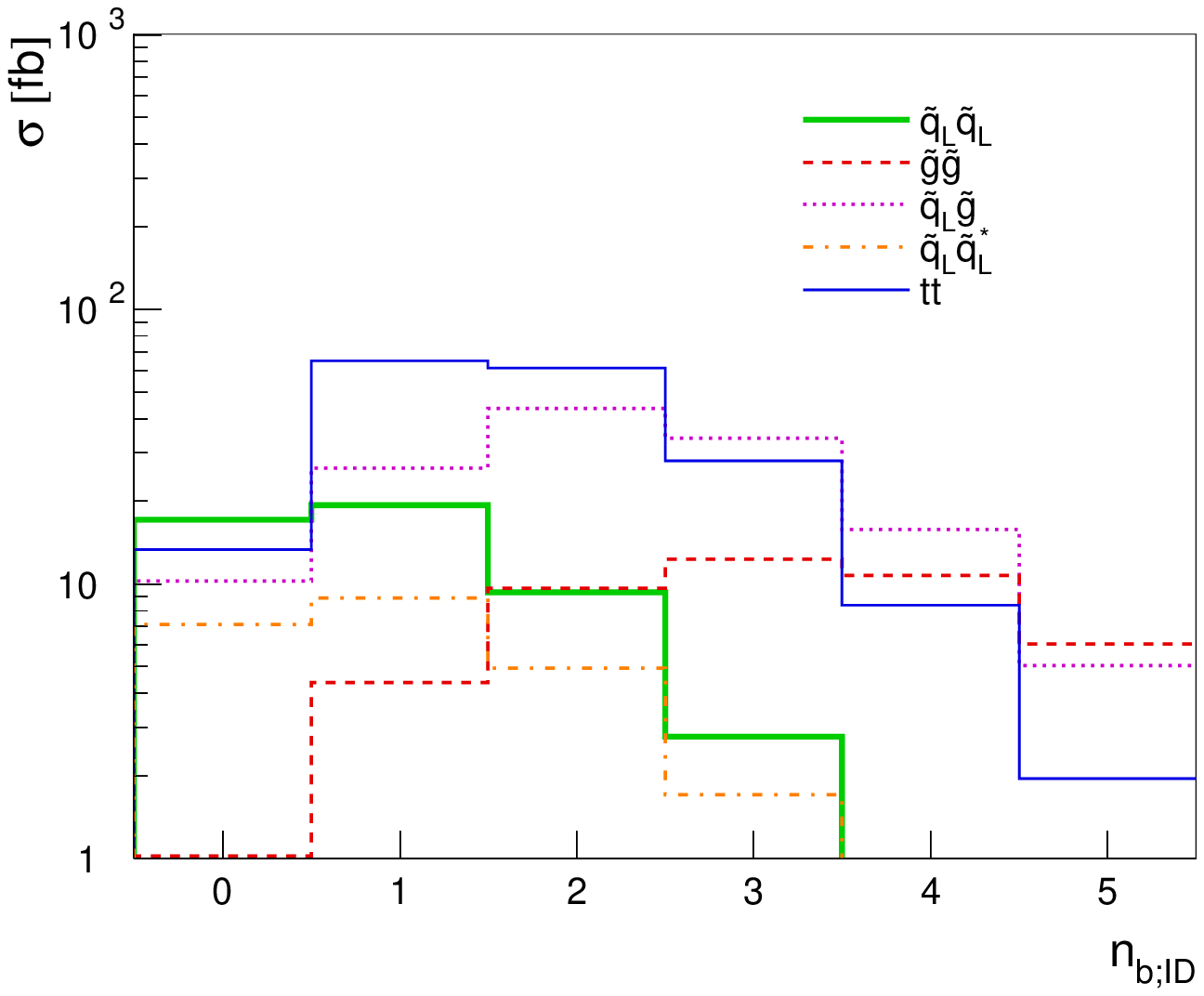}\vspace*{-4mm}\\
(a)  $\epsilon = 100$\% ; $D=0$\% & (b) $\epsilon = 90$\% ; $D=25$\% \\[-5mm]
\includegraphics*[scale=0.5]{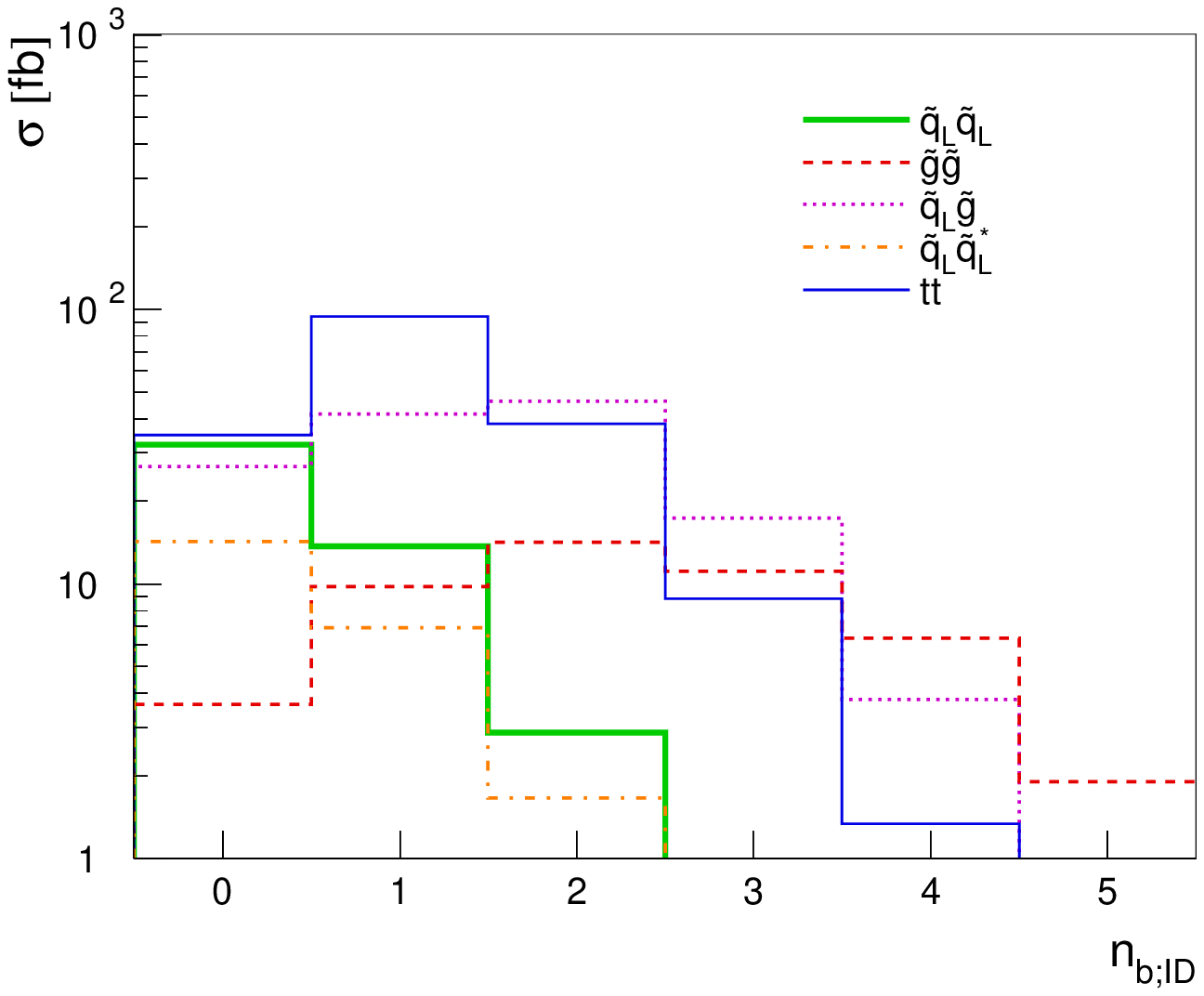}\vspace*{-4mm}&
\includegraphics*[scale=0.5]{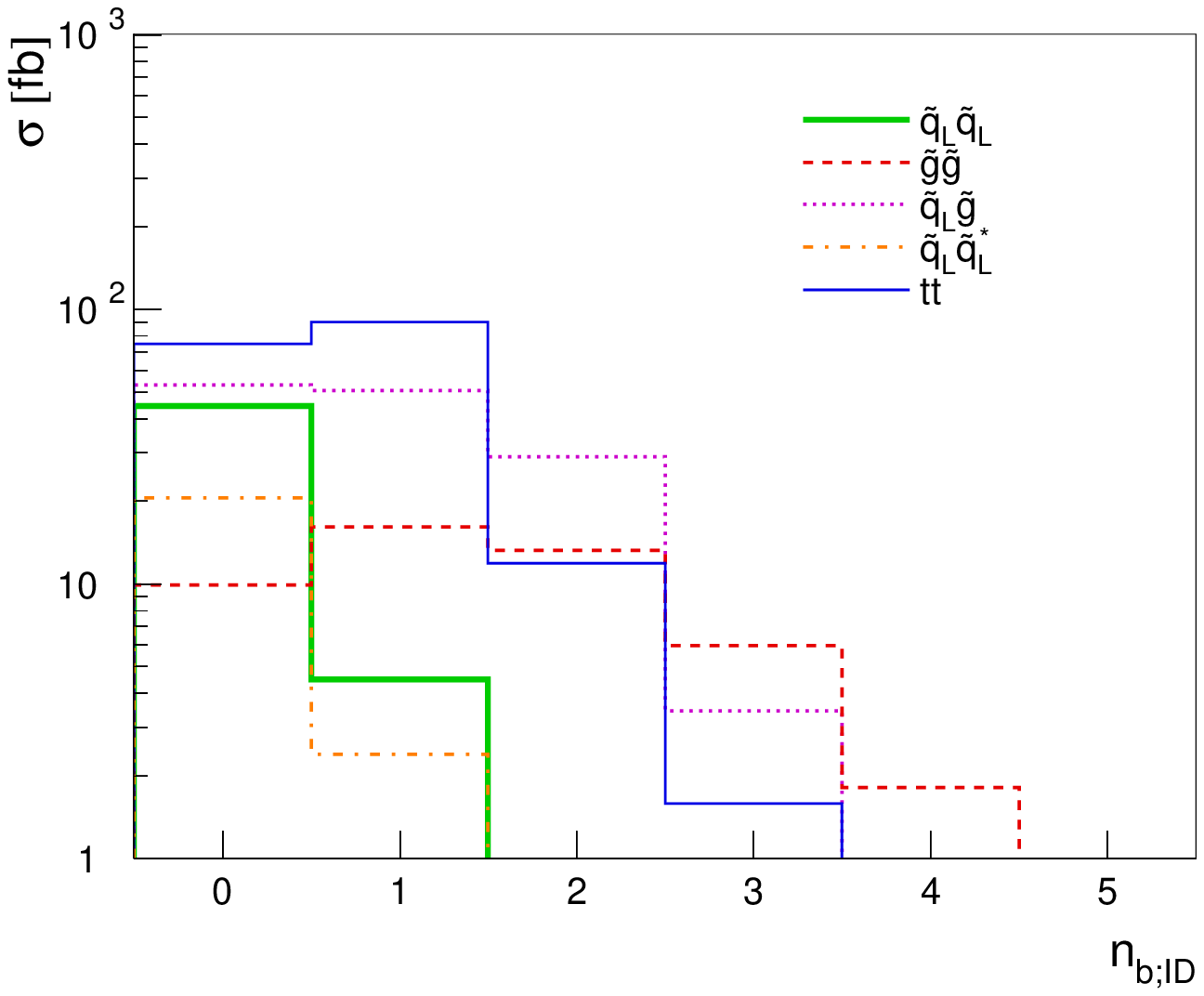}\vspace*{-4mm}\\
(c) $\epsilon = 80$\% ; $D=10$\% & (d) $\epsilon = 60$\% ; $D=2$\% 
\end{tabular}
\mycaption{Distribution of number of $b$ tags for the 4 different
  taggers discussed in the text, (a) being an idealized tagger and
  (b)-(d) ATLAS-inspired ones. 
\label{fg:nbj}} 
\end{figure}

The number of $b$ jets (inside $|\eta|<5$ and with $p_T>25$ GeV) which
would be identified with a perfect $b$ tagger 
is plotted in Fig.~\ref{fg:nbj}a. In reality, $b$
tagging is a trade-off between efficiency $\epsilon$ and mistagging
rate $D$. Following the ATLAS Physics TDR \cite{atlas} we
investigated the performance of 
three different b taggers; one, Fig.~\ref{fg:nbj}b, 
with a large efficiency $\epsilon = 90$\% making it efficient against
the background, but also high mistagging $D=25$\% rate, making it
expensive on the signal, a second, Fig.~\ref{fg:nbj}c, with $\epsilon = 80$\%
and $D=10$\%, and finally a third with  $\epsilon = 60$\% and
$D=2$\%. 

For our purposes, the most important goal is to obtain as pure a signal as
possible, and hence we choose the 0b bin of
the highest efficiency tagger, Fig.~\ref{fg:nbj}b, for this
analysis. For completeness, note that restricting the $b$ jet algorithm to the
central detector region $|\eta|<2.5$ would not greatly affect our
results; only a very small fraction of the true
$b$ jets lie outside this region. 

\begin{figure}
\vskip-4mm
\begin{tabular}{cc}
\includegraphics[scale=0.5]{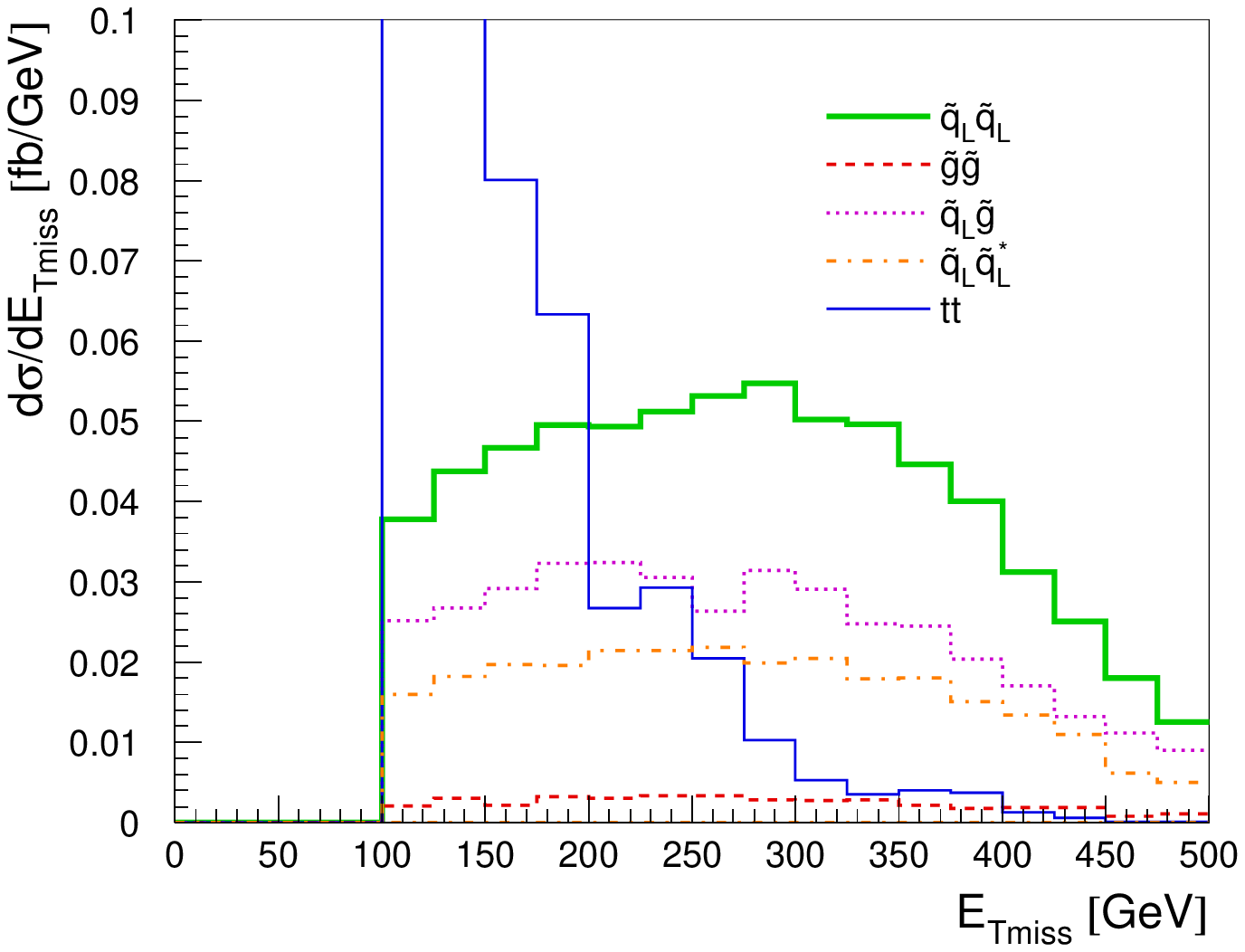}\vspace*{-4mm}
&
\includegraphics*[scale=0.5]{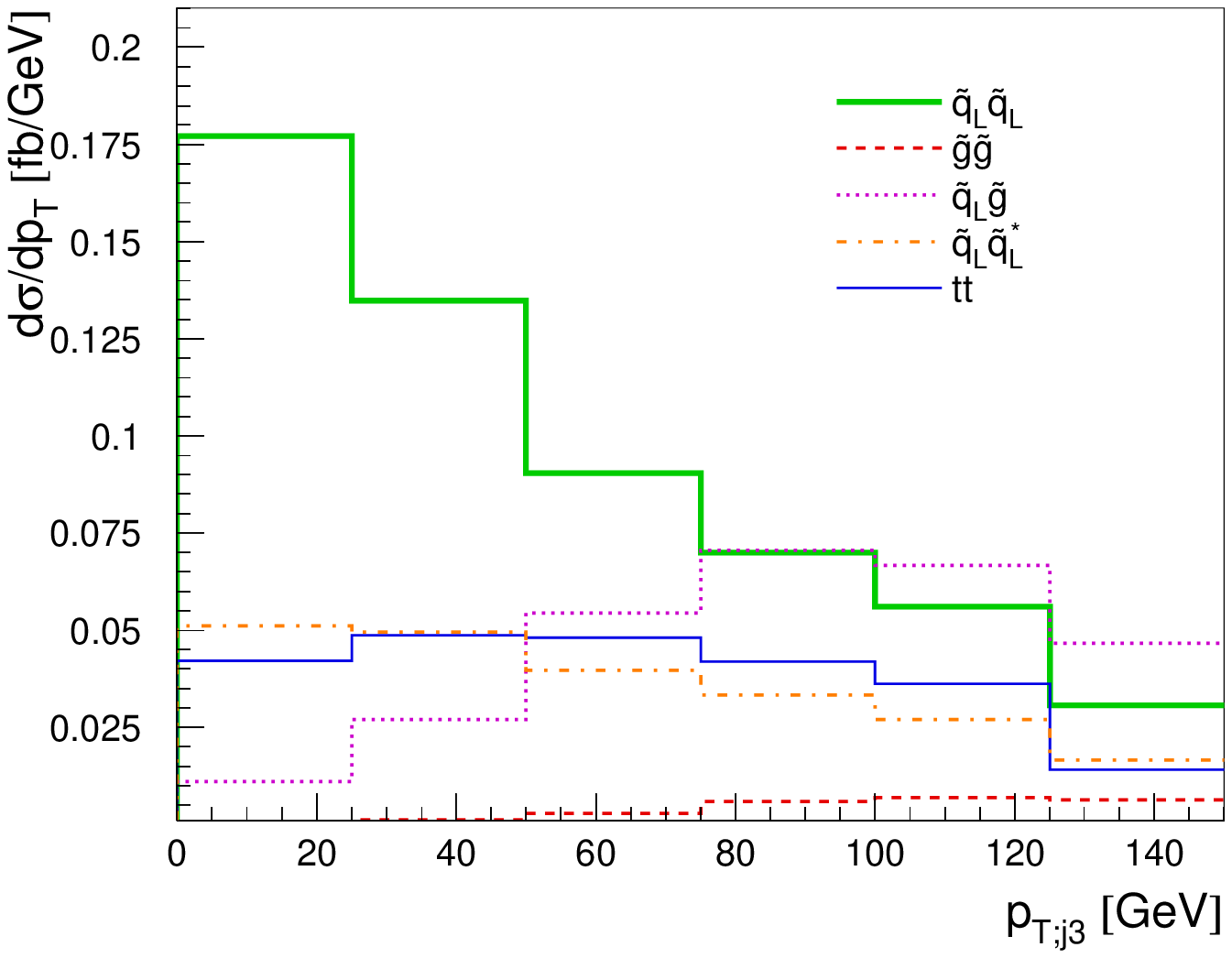}\vspace*{-4mm}\\
(a) & (b)
\end{tabular}
\mycaption{(a) Distribution of the
missing transverse energy after cuts on b-tagging. 
The bins below 100 GeV are cut out by the preselection.
(b) Distribution of
the transverse momentum of the third jet after the cut on $\Eslash$.}
\label{fig:eslash}
\end{figure}
Including now the $b$ jet veto, Fig.~\ref{fig:eslash} shows the
distribution of missing energy for the remaining events. 
While the contribution from top still
dominates the background it can be effectively suppressed by
increasing the $\Eslash$ cut. Here, we choose $\Eslash > 150$ GeV.

At this point, the gluino-related backgrounds dominate. 
A veto on hard additional jets is essentially
the only way of obtaining significant further rejection power against
these. Fig.~\ref{fig:eslash}~(b) shows the distribution of the transverse
momentum of the third-hardest jet for the events remaining in the
analysis. 
By rejecting all events with $p_{\rm T,j3} > 75$~GeV, the ratio of the
signal to gluino background is markedly improved. 

\begin{figure}
\vskip-4mm
\begin{tabular}{cc}
\includegraphics[scale=0.5]{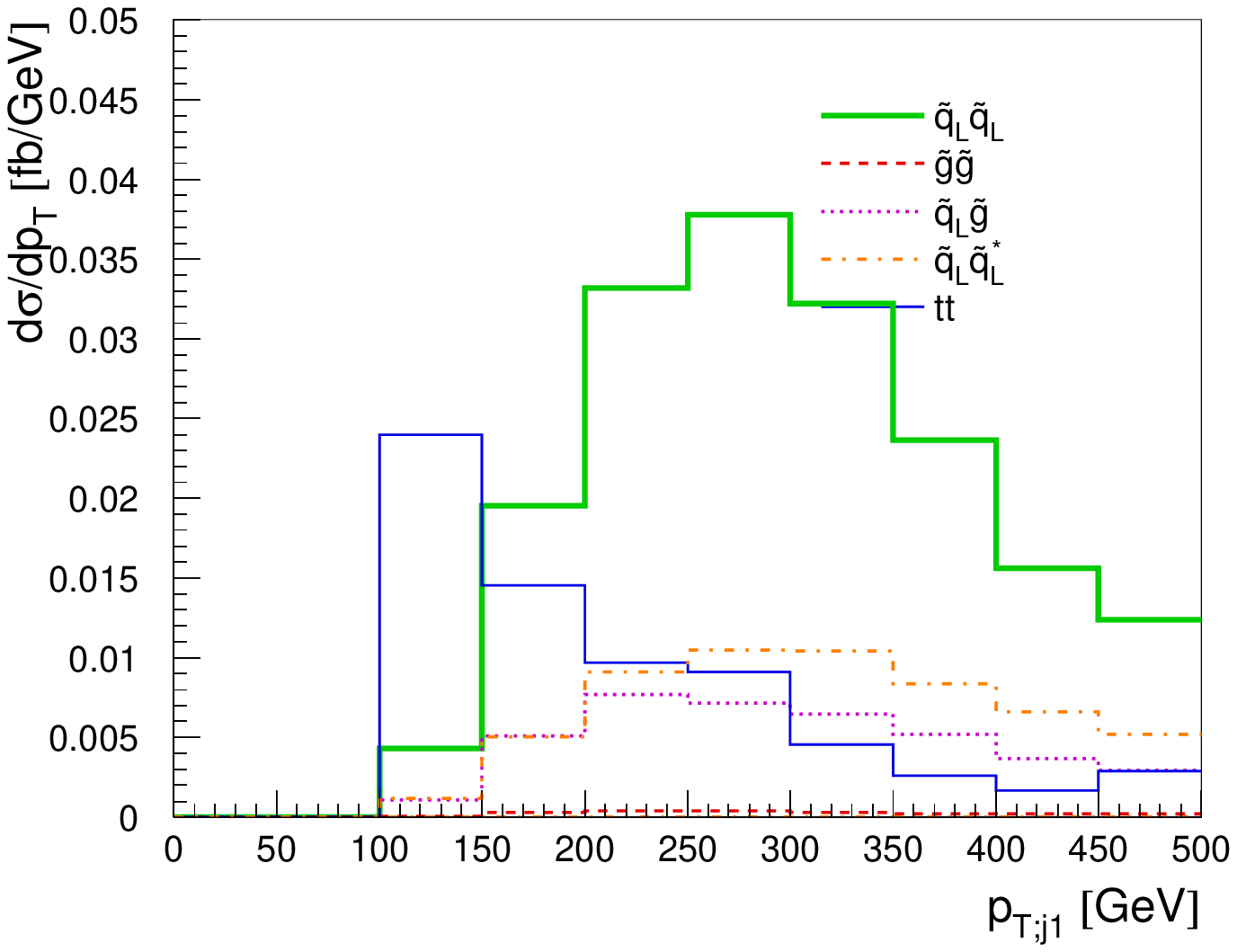}\vspace*{-4mm}
&
\includegraphics*[scale=0.5]{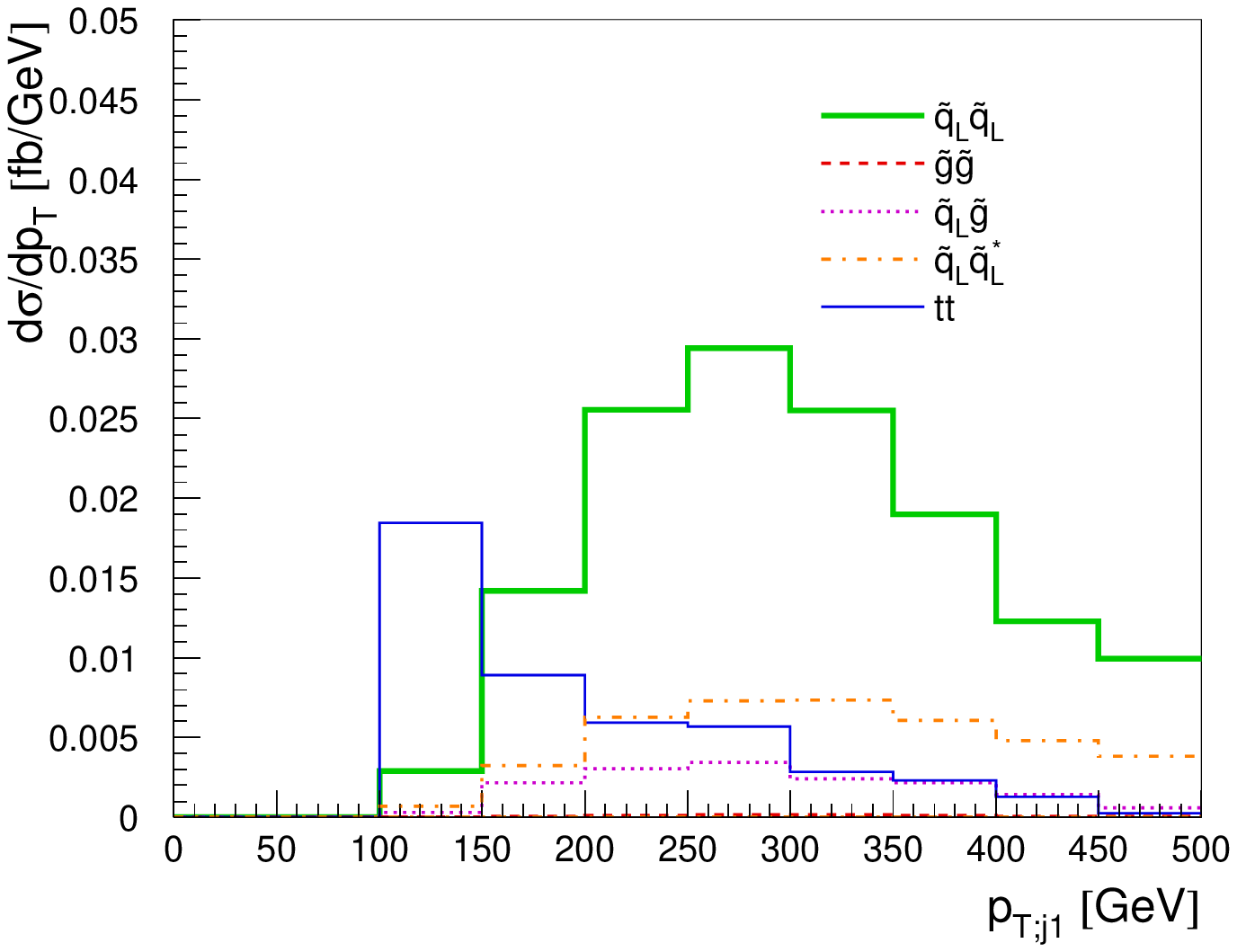}\vspace*{-4mm}\\
(a) & (b)
\end{tabular}
\mycaption{Distribution of the
transverse momentum of the hardest jet after the third jet veto for
(a) set A, and (b) set B.
The bins below 100 GeV are cut out by the preselection.}
\label{fig:pt1}
\end{figure}
The largest remaining background is now top production again (closely
followed by $\sqL\sqL^*$), whose two hard jets tend to be 
softer than those of the signal, due to the large
squark-chargino mass difference. This is illustrated in
Fig.~\ref{fig:pt1}~(a), where the transverse momentum of the hardest
jet in the remaining events is plotted. 

Increasing the transverse momentum cut on the first jet to $p_{\rm T,j1} >
200$ GeV we obtain a final signal cross section of 8.9~fb, over a
background of less than 7.6~fb, with the effect of progressive cuts on
signal and background normalisation and composition given in
Tab.~\ref{tab:cuts}. The signal-to-background ratio is 1.2, 
sufficient to allow a meaningful
measurement. With an integrated luminosity of 100~fb$^{-1}$, the statistical
error on the same-sign squark cross section is 4.5\%.

The reason we only quote an upper
estimate of the background in Tab.~\ref{tab:cuts} 
is that the $WWjj$ sample was not improved by
parton showering, hence we have no way of estimating the effects of
the 3rd jet veto on this component. 

\subsubsection{Harder Cuts --- Set B}

It is interesting to check  if the situation can be improved by choosing
stricter selection cuts. The gluino background can be further reduced by
tightening the cuts on the third jet to $p_{\rm T,j3} < 50$ GeV. In addition,
the signal is dominated by up-squarks, since two out of three valence quarks of
the proton are up-quarks. The $\tilde{q}\tilde{q}*$ background on the other
hand needs at least one sea-quark in the initial state and therefore results in
up- and down-squarks in more equal proportion. 
Thus the requirement of positively
charged leptons in the final state will increase the ratio of same-sign squarks
over opposite-sign squarks. With these stronger cuts the cross section values
on the last line of Tab.~\ref{tab:cuts} are obtained.  Indeed the
signal-to-background ratio is improved to 1.6, but at the cost of reduced
overall statistics. We choose a compromise; omitting the final cut on lepton
sign results in $S/B=1.45$, with acceptable  statistics, yielding a signal
cross section of 7~fb, over a total background of less than 4.9~fb. With a 100
fb$^{-1}$, the statistical error on the signal is 4.9\%. This error is slighty
larger than the result of the cuts Set A, but this is offset by lower
background systematics (see section~\ref{sc:comb}).

Noting that the two analyses are thus identical apart from the 3rd
jet veto, one might ask why it is important to show the results of
both. The reason, slightly touched on above and discussed
in more detail in \cite{susyjets}, is that 50 GeV jets are quite soft
objects at the LHC, 
especially as compared to the signal scales we consider here, of order
600-700 GeV. Since we cannot be entirely sure that the parton
shower resummation employed here is sufficiently accurate in this region, 
an explicit variation by a 25 GeV shift in the 3rd jet veto represents 
one way of verifying whether the analysis and
thereby our conclusions are stable against variations of this cut. 

\section{Squark decays at an \boldmath $e^+e^-$ collider}
\label{sc:ilc}
\subsection{Phenomenology and Strategy -- ILC}
In order to interpret the measured rates at the LHC in terms of the total squark
production cross section,
the individual branching ratios in the decay
chain eq.~\eqref{eq:dec} need to be established. We now turn to a
discussion of how each of 
these could be extracted from measurements at a high-energy
$e^+e^-$ linear collider.

The chargino properties are the most straightforward to determine,  
being easily accessible via the chargino pair production process,
$e^+e^- \to \cha_1^+ \cha^-_1$. Due to the large rate for that process, the
expected error on cross section measurements is about 1\% at a 500 GeV collider
\cite{lhclc}\footnote{Note that the SPS1a scenario studied in Ref.~\cite{lhclc}
is identical to the scenario in the appendix  in the electroweak sector.}.
Since the charginos $\cha^\pm_1$ are among the lightest supersymmetric
particles, the only open decay channels are $\cha^+_1 \to l^+ \nu_l \, \neu_1$,
$l = e,\mu,\tau$, which can all be easily separated from the background
\cite{lhclc}. We therefore assume that the chargino branching ratios are known
with 1\% error: BR($\cha^+_1 \to \tau^+ \nu_\tau \neu_1) = (100 \pm 1)$\%.

The squarks offer substantially more resistance. In 
the given scenario the left-squarks are in fact slightly too heavy to 
be accesible at a linear collider with 1 TeV center-of-mass
energy. Rather than attempting to find a more amenable parameter point, 
we here consider the hypothetical case of an $e^+e^-$ collider with
a center-of-mass energy of about 1.5 TeV. Obviously, we do not propose to address the technical challenges of such a machine, but 
justify our choice simply by the proof-of-concept nature of our study.
We also assume 
that both beams can be polarized, with polarization degrees
of 80\% for electrons, and 50\% for positrons. 

Having kept the squarks heavy, we must now deal with the fact that they 
can decay into the whole spectrum of charginos and neutralinos.
The heavier charginos and neutralinos have large branching ratios into
gauge bosons, 
which distinguishes them from the lighter states. The following characteristic
decay modes are chosen to distinguish the various charginos and neutralinos:
\begin{align}
\cha^+_1 &\to \tau^+ \nu_\tau \neu_1, & \mbox{BR=100\%} \label{eq:d1} \\[1ex]
\cha^+_2 &\to Z \cha^+_1 \to Z \, \tau^+\nu_\tau \, \neu_1, & \mbox{BR=24\%} \\[1ex]
\neu_2 &\to \tau\tau \, \neu_1, & \mbox{BR=100\%} \\[1ex]
\neu_{3,4} &\to W^\pm \cha^\mp_1 \to W^\pm \, \tau^\mp \nu_\tau \neu_1. &
\mbox{BR=59\%,52\%} \label{eq:d2}
\end{align}
A distinction of the $\neu_3$ and $\neu_4$ states is not necessary
since for the 
purpose of this study only the branching fraction into the lightest chargino
needs to be known as an absolute number. The tau leptons in the final state can
be identified in their hadronic decay mode with roughly 80\% tagging
efficiency. 
Since about 65\% of tau decays are hadronic, this amounts to a total tau
tagging efficiency of about 50\%, consistent with the findings e.g.\ 
of LEP2 studies
(see for instance Ref.~\cite{opalw}).

\subsection{Numerical Results -- ILC}
For this work, Monte-Carlo samples for squark pair production in the different
squark decay channels have been generated at the parton level with the tools of
Ref.~\cite{slep,blabla}. Also the most relevant backgrounds have been
simulated, stemming from 
double and triple gauge boson production as well as $t\bar{t}$
production. It is assumed that an integrated luminosity of 500 fb$^{-1}$ is
spent for a polarization combination $P(e^+)$/$P(e^-)$ = +50\%/$-$80\%, which
enhances the production cross section both for $\suL$ and $\sdL$ production.
Here $\mp$ indicates left/right-handed polarization.
The branching fractions are obtained from measuring the
cross sections of all accessible decay modes of the squarks and identifying the
fraction of decays into one specific decay mode out of these. 

Since the squarks are produced in charge-conjugated pairs, it is a priori
difficult to distinguish up- and down-squarks in the final state. However,
assuming universality between the first two generations, a separation between
up- and down-type squarks can be obtained through charm tagging. According to
Ref.~\cite{ilctag}, a $c$-tagging efficiency of 40\% is achievable for a purity
of 90\%. By combining the different decay channels in
eqs.~(\ref{eq:d1}--\ref{eq:d2}), the following final state signatures are
identified as interesting:
$jj(n\tau)\Eslash$ with $n\in\{1,2,3,4\}$, 
$cc(n\tau)\Eslash$ with $n\in\{2,4\}$, 
$jj\tau\tau (Z/W) \Eslash$, $cc\tau\tau Z\Eslash$,
where $j$ indicates an untagged jet, $c$ a tagged charm jet, and $Z/W$ a
hadronically decaying gauge boson where the invariant mass of the two jets
combines to the given gauge boson mass. 
For charged current squark decays, charm tagging does not provide any
additional information, since a
$cs$+leptons  final state can arise both from
up- and down-type squarks. 
Therefore, only signatures with two tagged charm jets are included in the list
above. Finally, since several squark decay channels can 
contribute to most of the final states above, 
one has to solve a linear equation system in order to derive
the individual contributions.

Before presenting results for this procedure, there is an additional
complication due to the subsequent decays of the neutralinos and charginos
after the squark decays. The branching fractions for the light chargino
$\cha^\pm_1$ and neutralino $\neu_2$ can be deduced from cross section
measurements at a low-energy run with $\sqrt{s} = 500$ GeV, as mentioned
above. Only a few decay channels are kinematically open, 
and the environment is sufficiently clean so that all decay
channels are easily visible with a small error.

The situation is more tedious for the heavier charginos and
neutralino, which have many different decay channels open. 
Rather than attempting to measure individual 
cross sections for all these decay modes, here instead a different method is
pursued. Near production threshold of two supersymmetric particles
$\tilde{X}\tilde{Y}$, the energy of a particle from a two-body decay mode of
e.g. $\tilde{X}$ is sharply defined. The signal of such a monoenergetic
particle would clearly stand out over the background, which is smooth in the
particle energy, and can thus be used to identify a
particular production cross section irrespective of the decay mode of the 
other sparticle $\tilde{Y}$. 

The branching ratios of the heavy neutralinos $\neu_{3,4}$ can be studied in
associated production with $\neu_2$, $e^+e^- \to \neu_2\neu_{3,4}$. When
approaching the production threshold, the decay of the $\neu_2$ is
characterized 
by a monoenergetic $\tau$ lepton over a smooth background. The production
threshold for $\neu_2\neu_3$ lies at $\sqrt{s} = \mneu{2}+\mneu{3}$, 
535 GeV in the present case. A 
few GeV above this threshold, at $\sqrt{s} = 540$ GeV, the production
cross section is already sizeable, $\sigma_{23}(\sqrt{s} = 540 \gev) = 15.8$ fb
with polarized beams $P(e^+)$/$P(e^-)$ = +50\%/$-$80\%. The cross section for
$\neu_2\neu_4$ is unfortunately 
too small to allow a meaningful measurement. An alternative is to 
obtain the branching fractions of $\neu_4$ at the $\neu_3\neu_4$ threshold,
by tagging the $\neu_3$ decays into a monoenergetic $W$-boson. Of course for
this purpose only hadronic $W$ modes can be used. A few GeV above the
kinematical threshold $\mneu{3}+\mneu{4} = 737$ GeV, the production
cross section is relatively large, $\sigma_{34}(\sqrt{s} = 745 \gev) =
29.3$ fb, assuming polarized beams as above.

For a full feasibility demonstration of this technique, 
a realistic detector simulation would be
necessary. Here the achievable precision for the neutralino and chargino
branching ratios is estimated by simulating the parton-level production near
threshold. Assuming 50 fb$^{-1}$ luminosity for each of the threshold
measurements, the following statistical errors are found:
\begin{equation}
\begin{gathered}
{\rm BR}(\neu_3 \to W^\pm \chi^\mp_1) = (59 \pm 6.5)\%, \qquad
{\rm BR}(\neu_4 \to W^\pm \chi^\mp_1) = (52 \pm 2.5)\%, \\
{\rm BR}(\cha^\pm_2 \to Z \chi^\pm_1) = (24 \pm 1.3)\%.
\end{gathered}
\end{equation}
Together with this information, the squark cross section measurements at
$\sqrt{s} = 1.5$ TeV can be interpreted in terms of squark branching ratios.
Solving the linear equations system that connects final state signatures with
production processes with a $\chi^2$ scan, the estimated errors for the squark
branching ratios are listed in Tab.~\ref{tab:sqbr}.
\begin{table}
\centering
\begin{tabular}{rrrr}
\hline
$\suL \to u \neu_1$ & $0.9 \pm 0.5$ \% &
 $\sdL \to d \neu_1$ & $1.9 \pm 0.8$ \% \\
$ u \neu_2$ & $29.0 \pm 3.0$ \% &
 $ d \neu_2$ & $28.3 \pm 4.8$ \% \\
$ u \neu_3$ & $< 1$ \% &
 $ d \neu_3$ & $< 0.2$ \% \\
$ u \neu_4$ & $< 1$ \% &
 $ d \neu_4$ & $1.9 \pm 0.8$ \% \\
$ d \:\! \cha^{\mbox{\tiny +}\!}_1$ & $67.7 \pm 3.2$ \% &
 $ u \cha^{\mbox{\tiny $-$}\!}_1$ & $63.9 \pm 5.2$ \%  \\
$ d \:\! \cha^{\mbox{\tiny +}\!}_2$ & $1.4 \pm 0.7$ \% &
 $ u \cha^{\mbox{\tiny $-$}\!}_2$ & $4.0 \pm 1.4$ \% \\
\hline
\end{tabular}
\mycaption{Estimated error for squark branching determination from measurements
at an $e^+e^-$ collider with $\sqrt{s} = 1.5$ TeV.}
\label{tab:sqbr}
\end{table}
For the purpose of this work, the interesting information are the branching
ratios of $\suL$ and $\sdL$ into the light chargino $\cha^\pm_1$.


\section{Combination -- LHC and ILC}
\label{sc:comb}

Based on the results from the simulations for squark production at the LHC and
the ILC presented in the previous sections, one can now derive an estimate for
the precision for the determination of the supersymmetric strong Yukawa
coupling $\hat{g}_{\rm s}$. Since the production of same-sign squarks at the
LHC with gluino $t$-channel exchange is proportional to
$\sigma[\sqL\sqL] \propto 
\hat{g}_{\rm s}^4$, the error for the coupling constant $\hat{g}_{\rm s}$ is
roughly one quarter of the error of the total same-sign squark production
cross section at the LHC. The statistical uncertainty is combined with the most
important systematic error sources in Tab.~\ref{tab:res}.
\begin{table}
\centering
\renewcommand{\arraystretch}{1.2}
\begin{tabular}{l@{\hspace{2cm}}rl}
\hline
 & $\sigma[\sqL\sqL]$ & $\hat{g}_{\rm s}/g_{\rm s}$ \\
\hline \hline
LHC signal statistics & 4.9\% & 1.3\% \\
SUSY-QCD Yukawa coupling & \\[-0.7ex]
\anc \ in $\sqL\go$ background & 2.4\% & 0.6\% \\
PDF uncertainty & 10\% & 2.4\% \\
NNLO corrections & 8\% & 2.0\% \\
Squark mass $\Delta m_{\sqL} = 9$ GeV & 6\% & 1.5\% \\
BR$[\sqL \to q' \, \cha^\pm_1]$ & 8.2\% & 2.0\% \\
\hline 
\hline
& 17.3\% & 4.1\% \\
\hline
\end{tabular}
\mycaption{Combination of statistical and systematic errors for the
determination of the absolute same-sign squark production
cross section at the LHC and the derivation of the strong SUSY-Yukawa coupling.
The cuts set B were used for the LHC.}
\label{tab:res}
\end{table}
The numbers are based on the choice (B) for the LHC signal selection in
Tab.~\ref{tab:cuts}. We now turn to a discussion of the various
unvertainties and caveats associated with these numbers.

The remaining background after selection cuts at the LHC introduces a
systematic 
uncertainty since the supersymmetric background from gluino production also
depends on $\hat{g}_{\rm s}$. Including the variation of $\hat{g}_{\rm s}$ in
the dominant squark-gluino background introduces an uncertainty of 5\% on the
signal. The background from opposite-sign squark production is also sizeable,
but depends in the same way on $\hat{g}_{\rm s}$ as the signal, $\sigma \propto
\hat{g}_{\rm s}^4$ and thus does not introduce an additional systematic effect.
 
In order to determine absolute cross sections, the proton parton distribution
functions (PDFs) in the relevant $x$ and $Q^2$ ranges must be known
accurately. The production of same-sign
squarks proceeds from same-sign quarks in the initial state and is therefore
dominated by the contributions from valence quarks. The value of valence
quark PDFs at high scattering energies can be computed reliably from
perturbative parton evolution and can be checked against measurements of vector
boson production. We tested this assumption by comparing results for different
CTEQ PDFs releases (CTEQ6M,  CTEQ6D,  CTEQ5M1, CTEQ5L and CTEQ6L1) \cite{cteq}
and find a maximal variation of 7\% in the signal and 4\% in the background. 
Thus we assign a total error of 10\% for the
parton distributions. 

Squark production receives large higher order QCD corrections. The
next-to-leading order corrections are known \cite{nlo} and can be included in
the analysis. The uncertainty of the missing ${\cal O}(\alpha^2_{\rm s})$ contributions
are estimated by varying the renormalization scale of the ${\cal
  O}(\alpha_{\rm s})$ 
corrected cross section between $m_{\sqL}/2 < Q < 2m_{\sqL}$, leading
to an error of 
8\%. Furthermore the cross section depends on the values of the squark masses.
According to Ref.~\cite{lhclc}, the left-chiral first generation squark masses
can be determined with an error better than $\Delta m_{\sqL} = 9$~GeV, leading
to an uncertainty of about 6\% for the production cross section. Finally, the
expected error for the determination of the squark branching ratios at the
linear collider must be included.

Combining all error sources in quadrature, it is found that the
same-sign squark 
production cross section can be determined with an error of 17.3\%,
translating to an error of 4.1\% for the 
supersymmetric QCD Yukawa coupling $\hat{g}_{\rm s}$. That is,
assuming the steps outlined here can be carried out in more or less
the fashion described, it should be possible to test the 
supersymmetry identity between gauge and Yukawa couplings 
to the level of a few percent.

It should be noted though that this result is highly scenario dependent. For
different supersymmetric scenarios, with larger squark masses, smaller gluino
masses or different squark decay modes, the analysis could be much more
difficult or even impossible. It is also interesting to study gluino production
not only as a background, but as a signal process for studying the SUSY-QCD
Yukawa coupling, although this introduces additional complications since gluino
production always depends both the the gauge and Yukawa
couplings. Nevertheless
we hope this study shows that the determination of the supersymmetric
QCD Yukawa coupling with competetive precision is not out of
reach, and that it is a tantalizing target for a complex 
all-embracing combination of LHC and lepton collider data. 

\section{Conclusion}
We have described a phenomenological method for testing
the identity between the QCD gauge coupling and the corresponding 
squark-gluino-quark Yukawa coupling that arises in
supersymmetric theories. Noting that same-sign
squark production at a hadron collider is both a very clean signal
and also directly 
proportional to the fourth power of the QCD Yukawa coupling, we
propose to measure the exclusive cross
section for same-sign squark production at the LHC in a particular decay
channel. To convert this to a measurement of the Yukawa coupling, 
branching fraction determinations at
a future lepton collider are (de-)convoluted with the LHC measurement 
to yield a combined determination of the total
inclusive cross section for same-sign squark production. We find that
this cross section can be determined to roughly 20\% precision, in the
scenario studied here, translating to roughly 5\% precision on the
Yukawa coupling. This is certainly encouraging and, we hope, 
sufficient in itself to spark off more in-depth studies. Obviously, 
due to the complexity and breadth of the proposed
analysis, we were forced to employ some shortcuts here, as described
in detail in the text above. It would be highly interesting to
establish how these estimates would compare to
both a more detailed (in the sense of detector capabilities and
systematic uncertainties) and comprehensive (in the sense of
exploring a wider range of parameter space) analysis. 

\section*{Appendix}

Here the reference scenario is listed that is used in this analysis. It is
coincides with the Snowmass point SPS1a \cite{sps}, but has a larger gluino mass
of 700 GeV. The scenario is defined at the weak scale through the supersymmetry
breaking parameters. The parameters relevant for this work are
\begin{equation}
\begin{aligned}
M_1 &= 99 \gev & \qquad\qquad m_{\rm L3} &= 197 \gev & \qquad\qquad m_{\rm Q1} &= 540 \gev \\
M_2 &= 193 \gev& m_{\rm R3} &= 136 \gev & m_{\rm U1} &= 522 \gev \\
M_3 &= 700 \gev& A_\tau &= -254 \gev & m_{\rm D1} &= 520 \gev \\
\mu &= 352 \gev& \tan\beta &= 10.
\end{aligned}
\end{equation}
The tree-level masses are computed from this to
\begin{equation}
\begin{aligned}
m_{\suL} &= 537 \gev & \qquad\qquad \mneu{1} &= 96 \gev \\
m_{\sdL} &= 543 \gev & \mneu{2} &= 177 \gev \\
m_{\st_1} &= 133 \gev & \mcha{1} &= 176 \gev \\
m_{\go} &= 700 \gev  & \mneu{3,4} &\sim 360 \gev. 
\end{aligned}
\end{equation}

\bigskip

\vspace{- .3 cm}
\section*{Acknowledgements}
We thank T.~Plehn for bug-free advice on Prospino.
A.F.\ is supported by the Schweizer Nationalfonds. P.S.\ is supported by
Universities Research Association Inc.\ under Contract No.\
DE-AC02-76CH03000 with the United States Department of Energy.


\end{document}